# From Fear to Hate: How the Covid-19 Pandemic Sparks Racial Animus in the United States

Runjing Lu and Yanying Sheng[1]

July 2, 2020

Click Here for the Latest Version

**Abstract**

We estimate the effect of the Coronavirus (Covid-19) pandemic on racial animus, as measured by Google searches and Twitter posts including a commonly used anti-Asian racial slur. Our empirical strategy exploits the plausibly exogenous variation in the timing of the first Covid-19 diagnosis across regions in the United States. We find that the first local diagnosis leads to an immediate increase in racist Google searches and Twitter posts, with the latter mainly coming from existing Twitter users posting the slur for the first time. This increase could indicate a rise in future hate crimes, as we document a strong correlation between the use of the slur and anti-Asian hate crimes using historic data. Moreover, we find that the rise in the animosity is directed at Asians rather than other minority groups and is stronger on days when the connection between the disease and Asians is more salient, as proxied by President Trump's tweets mentioning China and Covid-19 at the same time. In contrast, the negative economic impact of the pandemic plays little role in the initial increase in racial animus. Our results suggest that de-emphasizing the connection between the disease and a particular racial group can be effective in curbing current and future racial animus.

[1] Runjing Lu: University of Alberta (runjing1@ualberta.ca); Yanying Sheng: University of California San Diego (yas107@ucsd.edu). We are grateful for the support and guidance from Eli Berman, Julie Cullen, Gordon Dahl. We would also like to thank Ruixue Jia, William Mullins, Jacob Shapiro and participants at the UCSD applied lunch seminar for their constructive feedback.

# 1. Introduction

Racial animus can affect welfare in measurable ways, as economists have pointed out since the seminal work of Becker (1957). Recent papers have shown that racial animus can hinder economic development, affect political institutions, and induce social unrest.[2] To prevent further fueling of racial animus and reduce its damage on the society, it is fundamental to rigorously identify how racial animus forms and spreads.

The coronavirus (Covid-19) pandemic provides a natural experiment allowing us to study why and how racial animus against certain groups can rise rapidly. It has long been argued that infection risk can foster xenophobia (Schaller and Neuberg, 2012). While Covid-19 may originate from China, the Centers for Disease Control and Prevention (CDC) has emphasized that people of Asian descent are at no greater risk of spreading the virus than are other Americans. Nonetheless, since the outbreak of the virus, multiple incidents of Asian Americans being verbally or physically attacked have been reported by major news outlets.[3], Hundreds more have been recorded by organizations that track hate incidents (Mullis and Glenn, 2020). These are telltale signs that the current pandemic could have sparked racial animus against Asians. The challenge is to measure and understand this phenomenon rigorously and expediently.

In this paper, we exploit variation in the timing of the first local Covid-19 diagnosis across regions in the United States to causally identify how infectious diseases can trigger racial animus against Asians. We focus on the timing of the *first* Covid-19 diagnosis for two reasons. First, the first diagnosis in an area represents a salient increase in the infection risk. Lab experiments have shown that more salient threat of infectious diseases leads to stronger xenophobia (Faulkner et al., 2004). Second, the exact timing of the first diagnosis is plausibly exogenous --- whether an area has its first diagnosis this week or the next is largely unpredictable and is unlikely to be correlated with other factors simultaneously changing local racial animus.[4]

To proxy for an area's racial animus against Asians, we use the percentage of Google searches and Twitter posts (tweets) that include the word "chink" or "chinks". The proxy based on Google searches was first used by Stephens-Davidowitz (2014), who shows a negative relationship between an area's Google search rate of the word "nigger" or "niggers" and the vote share for Barack Obama in 2008, even after controlling for a number of demographic, economic and political variables. This proxy can capture *hidden* racial animus as searches are mostly private and unlikely to suffer from social-censoring. It has since been used to measure racial animus in papers like Anderson et al. (2020) and Depetris-Chauvin (2015) and been shown to have a positive relationship with economic downturn and anti-African American hate crimes (Anderson et al., 2020). To capture *public* display of racial animus, we use a proxy based on public tweets, which has been used by Nguyen et al. (2018) among others to measure sentiment towards minorities.

---

[2] For example, from the well-documented racial wage gap, residential racial segregation, costs of being minority political candidates, to the death of Gorge Floyd and the resulting protests across the United States --- these can all find their roots more or less in racism or racial animus (Charles and Guryan, 2008; Card et al., 2008; Stephens-Davidowitz, 2014; BBC, 2020).

[3] For example, *NBC News* (https://www.nbcnews.com/news/asian-america/video-shows-passenger-defending-asian-woman-facing-racism-new-york-n1162296), *New York Times* (https://nyti.ms/3ccvHzO), and *USA Today* (https://www.usatoday.com/story/news/politics/2020/05/20/coronavirus-hate-crimes-against-asian-americans-continue-rise/5212123002).

[4] Since the timing of the first diagnosis is generally earlier in areas with higher population size or better medical system, we include area fixed effects in all regressions to control for these time-invariant characteristics.

We focus on the use of the c-word, because this is the most salient racial slur against people of Asian descent and is unambiguously pejorative.[5] Indeed, Google searches and tweets including the epithet are mostly negative. "Chinked eye" and "chink virus" are common terms in racially charged Google queries, and over 53.4 percent of racially charged tweets are categorized as showing "anger" or "disgust" between November 2019 and April 2020. Moreover, an area's monthly Google searches for the epithet is positively correlated with monthly anti-Asian hate crimes between 2014 and 2018 and negatively correlated with monthly visits to Chinese restaurants between 2018 and 2019, even after controlling for area fixed effects and year-month fixed effects.

Our first main finding is that the Covid-19 pandemic leads to a surge in racial animus against Asians. In the week after the first local Covid-19 diagnosis, racially charged Google search rate increases by 22.6 percent of the area's maximum search rate over the sample period, and racially charged Twitter post rate increases by 100 percent of its average over the sample period. The result is robust to excluding early- and hard-hit states and to controlling for area fixed effects, year-month fixed effects, the severity of local Covid-19 infection, and an area's use of terms that are related to the c-word but are neutral in connotation, such as "Asian(s)", "Asia", or "China".

Analysis using unique Twitter user identifiers reveals that the increase in racially charged tweets mainly comes from existing Twitter users who post the racial epithet for the first time rather than those who have previously used the term. In the four weeks after the first local diagnosis, 2,064 Twitter users, who are not newly registered, tweet the epithet for the first time, potentially exposing their four million followers to racially charged content. This can create an multiplier effect on racial animus via persuasion (DellaVigna and Gentzkow, 2010) or by changing the social norms of publicly expressing anti-Asian sentiment (Bursztyn et al. 2020; Müller and Schwarz, 2019). Our findings also broadly relate to a growing literature on the role of social media in propagating animosity against minorities (Bursztyn et al. 2019; Müller and Schwarz, 2020).

Next, we turn our attention to the factors fueling racial animus. First, fear of infectious diseases could motivate racial animus. Evolutionary psychologists have argued that the desire to avoid harmful communicable diseases contributes to contemporary prejudices against subjective outgroups (Schaller and Neuberg, 2012). Surveys by Binder (2020) and Gallup (Saad, 2020) document that most Americans were worried about contracting Covid-19, suggesting that fear of the disease indeed exists in the United States. Second, the salience of the connection between Covid-19 and the Asian population is also a key factor. We find that the increase in animus is directed at Asians rather than other minority groups. Moreover, this racial animus is stronger on days when the connection is more salient, as proxied by President Trump's tweets mention China and Covid-19 at the same time. That time series relationship remains robust, even after we control for the severity of the pandemic and for general attention to Asians. Third, we find little evidence that the negative economic impact of the pandemic motivates the initial increase in racial animus. Areas bearing more severe economic impact of the pandemic do not exhibit a higher increase in racial animus than do those bearing less. This finding is consistent with surveys administered in early March and April 2020 which show that Americans are more worried about the effect of Covid-19 on their health than on their personal finances (Binder, 2020; Saad, 2020). Individuals may not fully comprehend the potential economic impact of Covid-19 at its onset.

---

[5] According to Philadelphia Bar Association (2014), the racial slur "originated in the 19th Century as a racial slur against people of Chinese descent", yet "is now widely used throughout the United States as a racial slur against people of Asian descent." The racial slur is as racist and hurtful to Asian Americans as the n-word is to African Americans (Richburg, 2008). Importantly, this racial slur has not been reclaimed by the Asian American community, as exemplified by the 2018 incident when TBS analyst Ron Darling, who himself is of partial Chinese descent, had to quickly issue a public apology after receiving criticism over his use of "chink in the armor" when referring to the performance of a Japanese pitcher playing for New York Yankees.

This paper builds on and contributes to the literature on the origin of racial animus. Past papers have documented that the deterioration of economic conditions can lead to animosity towards minorities. For example, Anderson et al. (2017) show that colder temperatures reduce agricultural production and intensify the persecution of Jewish people in Europe. Anderson et al. (2020) document that states hit harder by the Great Recession experience larger increases in racist Google searches and hate crimes against African Americans. In addition, evolutionary psychologists argue that fear of and desire to avoid health threats can motivate racial bias (Schaller and Neuberg, 2012). Earlier studies are mostly correlational based on surveys (e.g., Kim et al., 2016) or are established only in lab settings (Faulkner et al., 2004; O'Shea et al. 2020). We contribute by providing casual evidence that fear of infectious diseases and its link to a certain group lead to animus against the group while the economic impact of the disease plays a weaker role.

Our paper also contributes to the emerging literature on the relationship between Covid-19 and racial attitudes online and offline. Most papers are descriptive or correlational. For instance, Schild et al. (2020) characterize the evolution and emergence of Sinophobic slurs in the wake of the Covid-19 pandemic while Lyu et al. (2020) compares the characteristics of Twitter users who use versus do not use controversial terms when talking about the pandemic. One exception is the paper by Bartoš et al. (2020). They use a controlled money-burning task among subjects in the Czech Republic to show that elevating the salience of Covid-related thoughts magnifies hostility against foreigners living in Asia. We use a different empirical strategy and complement their paper by showing that infection risk gives rise to racial animus outside the lab as well. The fact that Covid-19 induces racial animus in both the United States and the Czech Republic suggests that the phenomenon documented in our papers is likely generalizable globally.

Finally, our work speaks to the literature on the role of political rhetoric. Political rhetoric has been shown to influence public opinions and behavior, such as presidential approval (Druckman and Homes, 2004) and public perception of a foreign country (Silver, 2016). In particular, Müller and Schwarz (2019) find that President Trump's tweets about Islam leads to anti-Muslim hate crimes. Our findings add to theirs by showing that the President's tweets also relate to anti-Asian sentiment, implying the generalizability of such a relationship to other racial attitudes.

Animosity between racial groups at an international level could severely hinder global initiatives to tackle the current pandemic and slow the economic recovery. Our results suggest that educating the public about the dissemination of Covid-19 and de-emphasizing the connection between the disease and a particular racial group can be an effective way to curb the current and future racial animus.

## 2. Data and Sample

*2.1. Google and Twitter Proxy for Racial Animus*

We use two measures to proxy for an area's racial animus against Asians --- the percentage of Google searches and Twitter posts that include the word "chink" or "chinks". The c-word is not uncommon in Google searches or in tweets. Between June 2019 and June 2020, the racial epithet was included in over a quarter million Google searches and 60 thousand tweets.[6] This is about the same number of Google searches as the word "democrat" and 3 percent of tweets as the word "economists".

We obtain the data from two sources. First, Google search data are obtained using Google Trends. We download weekly Google search data for the c-word at the media market level between July 2019 and April 2020. The data are not the raw number of searches, but the weekly percentage of searches including the term over all searches in a media market *search rate*, scaled by the highest weekly search rate in the same market during the whole time period that one extracts the data for --- in our case, between July 2019 and April 2020. In particular, racially charged Google search index for media market *m* at time *t* extracted over period *T* is approximately:

$$\text{Search Index}_{mt,T} = 100 \times \frac{\frac{\text{Searches including "chink(s)"}_{mt}}{\text{Total searches}_{mt}}}{\max_{t \in T}\{\frac{\text{Searches including "chink(s)"}_{mt}}{\text{Total searches}_{mt}}\}} \quad (1)$$

This metric is able to capture the timing of a change in an area's racially charged search index but not the absolute level of the change. In Appendix A, we re-scale the search index so that the index in different media markets is normalized using the same base. However, doing so leaves us with only 60 percent of the original sample. Therefore, we focus on the timing of the change in the main results and report the results using the re-scaled version in the appendix.

---

[6] The number of Google searches are approximations from https://searchvolume.io/, a free of charge substitute to Google AdWords. The data are only available for the 12-month period before our query on June 8th, 2020.

We are able to extract racially charged Google search index for 60 over 210 media markets, covering around 40 percent of the U.S. population in 33 states. We are not able to extract the data for other media markets because Google does not report the data when the search index for a given area and time is below an unreported threshold. Compared to media markets with no racially charged search index, those in our sample have higher population and exhibit a quadratic relationship with the percentage of Asian population but do not differ in other measurable dimensions, such as local unemployment rate or support for democratic versus republican party (Table A1).[7] Analyses using Google data are done at the media market level.

Second, Twitter data are obtained from Crimson Hexagon, which houses all public tweets through a direct partnership with Twitter. We download all geo-located tweets that include the c-word between November 2019 and April 2020. Crimson Hexagon does not provide data on the total number of tweets posted in a given area and time. We thus extract the number of all public tweets including "the", the most common word on Twitter, to approximate the total Twitter activity in each county on each day. The assumption is that the proportion of tweets including "the" is stable across areas, and the number of tweets including "the" can approximate total activity on Twitter. We define racially charged Twitter post index for a given county and time as the number of tweets that include the c-word per 100,000 tweets including "the".

We have valid Twitter post index in 612 counties across 51 states, encompassing roughly 59.5 percent of the U.S. population. Counties are not in the sample because their residents do not use Twitter, do not disclose geo-identifiers on Twitter, or do not post any tweets including "the" in the sample period.[8] Counties with valid Twitter post index on average have a larger population, more educated residents, and slightly higher support for democratic party than do counties without the data (Table A1). Analyses using Twitter data are done at the county level unless noted otherwise.

Admittedly, Google and Twitter data suffer from sample selection either due to low search activities or missing geo-identifiers. But given that areas with Google data and those with Twitter data are not highly correlated (correlation=0.053 at county level), using both data sources can alleviate the concern about the external validity of our findings.

*2.2. Relationship between Racial Animus, Hate Crimes, and Consumer Decisions*

For racially charged Google search index and Twitter post index to be meaningful proxies for racial animus, the only assumption we need is that an increase in racial animus makes one more likely to use the c-word. Under this assumption, higher racial animus will result in a higher percentage of Google searches and tweets that include the racial epithet. Existing papers that use a similar proxy for racial animus suggest that the assumption is likely to hold in reality (Anderson et al., 2020; Depetris-Chauvin, 2015; Stephens-Davidowitz, 2014). Common terms in racially charged Google searches and tweets also support the assumption. During our sample period, some most common terms in these searches are "chinked eye" and "chink virus"; common terms in these tweets are phrases like "chink virus", or directly addressing another individual as a "chink". Furthermore, over 67 percent of these tweets are tagged with emotion of "anger" or "disgust".[9]

---

[7] The quadratic relationship between the percentage of Asian population and racial animus is consistent with the theory of racial threat (Glaser, 1994). In communities with zero Asians, race is not salient and racial animus is less likely to form. In communities where Asians account for almost 100 percent, there are very few white individuals and those with racial animus are unlikely to choose such a community.

[8] Geo-identifier is voluntarily provided when users sign up for Twitter. Roughly 50 percent of tweets have valid geo-codes in our sample.

We benchmark our proxies with common measurement of racial animus and consumer discrimination. We begin by presenting the relationship between the proxies and anti-Asian hate crimes. Hate crime data come from the FBI Uniform Crime Reports (UCR) Hate Crime Statistics and are available up to 2018.[10] A majority of these hate crimes are personal crimes, including simple or aggravated assault (30 percent) and in-person intimidation (34 percent). Table 1 panel A columns (1) through (4) report the correlation between the monthly racially charged search index and monthly anti-Asian hate crime rate in each media market between January 2014 and December 2018. On average, a one standard deviation increase in racially charged search index (i.e., 29.6) link to around 15 percent $(= 29.6 \times 0.00019/0.03746)$ increase in the average monthly anti-Asian hate crime rate in each media market each month, controlling for media market and year-month fixed effects. The correlation is robust to controlling for the search index for "Asian(s)", which are related to the c-word but neutral in connotation. The relationship between racially charged search index and hate crimes is mainly contemporaneous as the coefficient on the last month's search index is small and insignificant.

Next, we change the dependent variable to monthly visits to Chinese restaurants in the local area between January 2018 and December 2019. Restaurant visit data are from Safegraph and available starting in 2018.[11] As shown in Table 1 panel A columns (7) and (8), a one standard deviation increase in racially charged search index links to roughly 160 to 190 fewer visits to Chinese restaurants per one million population in each media market each month, controlling for media market and year month fixed effects as well as search index for "Asian(s)". This decrease equals to about 0.6 to 0.7 percent of the average monthly visit rate. The relationship between search index and visit rate is also mainly contemporaneous.

We replicate the above correlations using Twitter data in Table 1 panel B. We aggregate hate crimes to the media market level due to low occurrence at the county level. To maintain consistency, we also aggregate restaurant visits to the media market level. Overall, racially charged Twitter post index does not correlate with anti-Asian hate crimes or visits to Chinese restaurants. One potential explanation is that Twitter data represent the public display of racial animus and undergo higher social censoring. We may only see a change on Twitter when the shift in racial animus is substantially large.

---

[9] Crimson Hexagon assigns each tweet one or more emotion tag(s) generated via natural language processing algorithm. The algorithms can distinguish the difference in emotions between tweets even when the words included are similar, for instance, the difference between "I want a burrito so bad" and "I just had a burrito. It was so bad.". For a detailed description, please refer to https://www.brandwatch.com/blog/understanding-sentiment-analysis.

[10] Over 80 percent of the U.S. population is covered by police agencies that voluntarily report hate crime data to UCR (Ryan and Leeson, 2011).

[11] Safegraph partners with mobile applications and collects anonymous user location data to calculate the foot traffic to around 4.1 million points of interest in the United States.

*2.3. The Covid-19 Pandemic in the United States*

Our empirical strategy relies on the plausibly exogenous variation in the timing of the first Covid-19 diagnosis across regions in the United States. We download the data on all Covid-19 cases and deaths in the United States between Jan 21st and April 26th, 2020 from John Hopkins University Covronavirus Resource Center. We match the date of the first Covid-19 diagnosis in each county and media market to those with valid Google and Twitter data. All media markets have their first diagnosis in the sample period and have at least six weeks after the first diagnosis. We exclude seven counties that do not have diagnoses in the sample period and 18 counties whose first diagnosis is after March 29th 2020 to ensure that all counties have at least four weeks after the first diagnosis. The remaining counties and media markets constitute our main regression sample. The county- and media-market-level predictors of being in the main sample are presented in Table A1. Figure A1 plots the location of the regions included in the main sample by the timing of their first Covid-19 diagnoses. Overall, coastal and large metropolitan areas, such as Orange county in California and Cook county in Illinois, are the earliest hit while those in the middle of the country, such as Adams county in Colorado, are among the last hit. We formally show the relationship between local characteristics and the timing of first local Covid-19 diagnosis in Table A2. Areas with higher population and more males tend to have their first cases earlier but, interestingly, areas with more Asians do not.

In Figure 1 panel A, we plot the number of counties and media markets by the week of their first Covid-19 diagnoses. The week of the first diagnosis is as early as the week of January 19 2020 and as late as the week of March 29 2020.[12] To visually gauge the relationship between the timing of the first diagnosis and the use of the c-word on Google and Twitter, we plot the U.S. weekly racially charged Google search index and Twitter post index in Figure 1 panel B. The sharp rise in the search index around early March and the rise in post index around mid-March correspond well with the waves of first Covid-19 diagnoses at the media market and the county level.

---

[12] There is more variation in the timing of the first Covid-19 diagnosis measured in day, as shown in Figure A2. The daily search and the daily post index are much noisier than the weekly version. We thus present the weekly version as the main results and the daily version as robustness check in the appendix.

## 3. Strategy and Results

*3.1 Empirical Strategy*

Our main strategy is a difference-in-differences event study where the first Covid-19 diagnosis in a county or a media market is the event of interest. The specification is as follows:

$$Y_{it} = \sum_{k=-6}^{-2} \beta_k \times 1\{k = t - E_i\} + \sum_{k=0}^{4 \text{ or } 6} \beta_k \times 1\{k = t - E_i\} + \gamma' X_{it} + \alpha_i + \alpha_{ym} + \epsilon_{it}, \quad (2)$$

where $Y_{it}$ is racially charged Google search index or Twitter post index in county or media market $i$ in week $t$. $E_i$ is the week when $i$ has its first Covid-19 diagnosis. $1\{k = t - E_i\}$ refers to event dummies indicating the number of weeks before or after the first local diagnosis. We omit the dummy for the week before the first diagnosis due to perfect collinearity. We include six weeks of post-period for Google data and four weeks for Twitter data because longer post-period will result in unbalanced sample in event time. $X_{it}$ is a vector of area-specific controls such as the number of diagnoses or deaths related to Covid-19, an indicator for state-level stay-at-home order, and search index or post index for "Asian(s)". We include area fixed effects ($\alpha_i$) and year-month fixed effects ($\alpha_{ym}$) to control for an area's average level of racially charged search index and post index as well as the national trend in the indices.[13] $\epsilon_{it}$ is the error term. Standard errors are clustered by media market for Google data and county for Twitter data. For a media market that crosses state border, we assign it the state where the highest fraction of its population reside. To understand how immediately the first local Covid-19 diagnosis has an effect on local racial animus, we run regression 2 at the daily level and additionally control for day-of-week fixed effects.

The coefficients of interest are $\beta_k$'s when $k \geq 0$, which indicate the dynamic effect of an area's first Covid-19 diagnosis on racially charged search index and post index. The identifying assumption is that the progression of racially charged search index and post index in areas that have and not yet have the first Covid-19 diagnosis share parallel trends in the absence of the Covid-19 pandemic. This assumption is inherently untestable, but we can assess its plausibility by testing for parallel pre-trends. We provide evidence that the assumption is likely to hold in the next section.

*3.2. The Effects of Covid-19 on Local Racial Animus*

We start by examining how an area's search of the c-word on Google respond to the first local Covid-19 diagnosis. Figure 2 panel A plots the estimates of the coefficients on the event dummies from regression 2 using an area's racially charged Google search index as the outcome. The search index reaches the peak in the week after the first local diagnosis and decreases slightly in later weeks. Table A3 panel A report the regression results. Given the construction of the search index, the estimates should be interpreted as a percentage of an area's maximum search rate in the sample period. Therefore, compared to the week before the first Covid-19 diagnosis, racially charged search rate increases by 22.6 percent of the area's maximum search rate over the sample period in the first week after the first diagnosis and remains at least 15 percent till six weeks afterward.[14] Given what we find in Table 1, the increase in search index in the four weeks after the first Covid-19 diagnosis corresponds to an increase of 0.0033 ($0.0002 \times (22.63 + 16.95 + 8.16 + 19)/4$) anti-Asian hate crimes per million residents, or 10 percent of the average monthly anti-Asian hate crime rate.

---

[13] Although the Google search rate is normalized so that the maximum rate is 100 for each area, there is still considerable variation in the sample mean, i.e.., the average *actual* search rate over the *actual* maximum rate, ranging between 8 and 50.

We also present the estimates using the search index scaled by the same maximum of Fresno-Visalia in California in Figure A3 as a robustness check. This is because the estimates using the original search index do not map to an increase over a national base, as the maximum search rate is different across areas. The overall pattern is similar whether we use the rescaled or the original index, but the magnitude of the former is about one-third to two-thirds of the latter because the maximum of Fresno-Visalia is larger than that of most media markets. The confidence intervals are also larger because 40 percent of the sample are lost when we re-scale.[15]

To better interpret the effects and understand how public expression of the c-word changes, we turn our attention to Twitter. Similar to Google search index, racially charged Twitter index also peaks in the week after the first local Covid-19 diagnosis and slowly decreases afterwards, as plotted in Figure 2 panel B. Table A4 panel B further shows that, relative to the week before the first diagnosis, racially charged tweets increase by 1 and 0.6 per 100,000 "the" tweets in the first and second week after the first diagnosis. The increase amounts to roughly 100 and 66 percent of the outcome mean. In columns (2) through (4) in in Table A3 and A4, we control for the number of confirmed cases and deaths, whether a state-level stay-at-home order is in place, general interest in terms like "Asian(s)", and excluding early- and hard-hit state, such as New York, Washington, and California. The estimated effects increase slightly after controlling for the covariates.

To get at the exact timing when the usage of the c-word starts to change, we run regression 2 using the daily search and post index as outcomes, additionally controlling for day-of-week fixed effects. Regression estimates are plotted in Figure A4. The effects start to appear around two to three days after the first local diagnosis. The local residents indeed pay attention to the Covid-19 situation in their neighborhood and respond quickly to the first diagnosis.

One may concern that the increase in the usage of the c-word is driven by reasons other than higher animosity against Asians, such as an increase in online activities due to blanket stay-at-home orders, a rise in benign attention to China or Asia, seasonality in racist online activities, or "Twitter bots". These factors are unlikely to explain our findings. First, the search and the post index are normalized by the total searches or tweets in a given area and time and thus already account for overall change in online activities. Second, Table A3 column (3) shows that our results are robust to controlling for the search and the post index for terms which capture general attention to China or Asia but are neutral in connotation, such as "Asian(s)", "China", or "Asia".[16] Third, to test the seasonality in racist online activities, we generate a "fake" Covid-19 diagnosis date for each area using the same calendar day and month of its actual diagnosis date but change the year to 2019. Reassuringly, we find no increase in racially charged search and post index surrounding the "fake" dates, as shown in Figure A5. Finally, some may worry that "Twitter bots" rather than local residents contribute to the increase in racially charged tweets. Twitter proactively identifies and removes automation generated content which can alleviate this concern to some extent (Roth and Pickles, 2020). In addition, our results do not quantitatively change when we drop frequent c-word users who are subject to suspicion of to be bots, defined as those who used the epithet more than five times (99 percentile) during the sample period. Results are available upon request.

---

[14] The increase during the week of the diagnosis is small because the diagnosis may occur late in the week.
[15] The detailed re-scaling procedure is in Appendix A.

So far, we have shown that the Covid-19 pandemic raises both individuals' *hidden* animosity against Asians and their *public* display of the animosity. A natural next question is who contributes to the rise. Do more individuals start to harbor the animus or do a few existing racists increase their animosity? Taking advantage of unique Twitter user identifiers, we can breakdown the increase in racially charged tweets by whether their authors are first-time or existing c-word users. We define *existing* c-word users as those who tweeted the c-word at least once between 2014 and the sixth week before the first local Covid-19 diagnosis. We define *new* c-word users as those who never tweeted the c-word between 2014 and the sixth week before the first local diagnosis and who had at least 10 tweets before their first c-word tweets. Importantly, this definition can avoid counting newly registered Twitter users as new c-word users. Figure 3 plots the breakdown. The increase in racially charged tweets is mainly driven by new c-word users rather than pre-existing ones. This breakdown suggests that the Covid-19 pandemic induces *more* existing Twitter users to start publicly expressing animus against Asians. To be specific, in the four weeks after the first local diagnosis, 2,064 Twitter users started to use the racial epithet, which could expose their four million followers to racially charged content induced by the Covid-19 pandemic. This can create an multiplier effect by persuading more individuals to hold such racial attitudes via an increase in exposure to anti-Asian sentiment (DellaVigna and Gentzkow, 2010) or by changing the social norms of publicly using the racial epithet (Bursztyn et al. 2020; Müller and Schwarz, 2019).

---

[16] We only tabulate results controlling for "Asian(s)". Results controlling for other relevant terms are quantitatively similar and available upon request.

*3.3. Factors Fueling Local Racial Animus*

In this section, we discuss several non-mutuality exclusive explanations as why the Covid-19 pandemic spurs animosity against Asians.

**Fear of infectious diseases.** Evolutionary psychologists have long argued that desire to avoid harmful communicable diseases contributes to contemporary prejudices against subjective outgroups (Schaller and Neuberg, 2012). Lab experiments also show that xenophobia toward unfamiliar immigrant groups is stronger when the threat of infectious diseases is more salient (Faulkner et al., 2004). Moreover, surveys administered in March and April 2020 document that around 60 to 80 percent of Americans are worried about contracting Covid-19, suggesting that fear of the disease indeed exists among the locals (Binder, 2020; Saad, 2020). In our main analysis, we also show that local racial animus responds to a salient increases of infection risk.

**Connection between Covid-19 and Asians.** The salience of the connection between Covid-19 and the Asian population could also play a key role in fueling the animosity against Asians. First, if the salient connection is not a main driving force, the disease-avoidance theory would predict rising animus against all minorities, not just Asians. We construct Google search index and Twitter post index for common racial epithets against major minority groups in the United States, such as "nigger(s)" against African Americans, "wetback(s)" against Hispanics, and "kike(s)" against the Jewish population.[17] We run regression 2 using racially charged searches and tweets against other minorities as outcomes. We include an indicator for the week of January 26 2020 when using the n-word as the outcome, because Kobe Bryant's death together with MSNBC Anchor using the n-word when broadcasting the news of the death leads to a spike in its use. We also include an indicator for the week of February 23 2020 when using the k-word as the outcome, because Los Angeles Dodges player Enrique ("Kiké") Hernandez's performance in that week leads to spike in the use of the k-word. Coefficients on the event dummies are plotted in Figure A6. None of the examined racial epithets experience increase in Google searches following the first local diagnosis; if anything, searches for the n-word see a slight decrease. Similar pattern is found for the w-word and the k-word on Twitter.[18] These findings suggest that the *connection*

---

[17] We do not use "spic(s)" as the racial epithet against Hispanics, because the term is included in "Spic and Span", the name of an all-purpose household cleaner brand, which experienced growing interest since the start of the Covid-19 outbreak. Breakout Google queries and a substantial number of tweets including the term are about the brand and not the slur. In addition, we do not include "redskin(s)", a common racial epithet against Native Americans, because the term is included in the name of a professional American football team "The Washington Redskins". Google queries and tweets including the term are mostly about the football team, such as, "chase young redskins" and "redskins draft".

[18] We present Twitter post index for the n-word separately in Figure A7 due to the seasonality in the use of the n-word on Twitter. The seasonality is evident when comparing the n-word usage between 2019 and 2020 in panel A. The seasonality may arise from a combination of Black History Month taking place in February and the n-word being reclaimed by African Americans (Croom 2011). These factors may invalidate the use of the term on Twitter as a proxy for racial animus. Note that we additionally include an indicator for the week of February 9 2020 in panel A,

between Covid-19 and the Asian population, not just fear of contracting Covid-19 from unfamiliar outgroups, drives the rising animus against Asians.

Second, we can examine how racial animus against Asians varies with the salience of the connection between Covid-19 and the Asian population. We proxy for the salience of this connection using President Trump's tweets mentioning Covid-19 and China at the same time.[19] President Trump has 82.4 million followers on Twitter, and his tweets have been shown to affect public behavior like hate crimes (Müller and Schwarz, 2019). We expect to see higher racially charged searches and tweets on days when the connection is more salient. This is exactly what we find. Table 2 columns (1) through (3) show that there are five more racially charged tweet per million "the" tweets in a day when President Trump mentions China and Covid-19 together in roughly six more tweets ($= 5/(0.0829 \times 10)$). The increase amounts to 12 percent of the average rate of racially charged tweets ($= 0.0829/0.669$). Columns (4) through (6) display a similar but weaker relationship between President Trump's tweets and racially charged searches on Google. The weaker relationship may be because not all Google users have Twitter accounts and may not respond to viral events on Twitter.

Importantly, the rate of racially charged tweets and searches do not correlate with the President's tweets mentioning *only* China or *only* Covid-19, controlling for county and time fixed effects. Moreover, the findings remain quantitatively similar after we control for the daily number of new Covid-19 diagnoses and deaths in the United States and the daily rate of tweets about "Asian(s)". Therefore, an increase in the severity of Covid-19 pandemic or benign attention to the racial group could not explain our findings. In sum, the salience of the connection between the disease and Asians propagates racial animus in the current pandemic.

**Economic Downturn.** Researchers have documented that the deterioration of economic conditions can fuel animus towards minorities (Sharma, 2015; Anderson et al., 2017, 2018). The Covid-19 pandemic imposes risks on both lives and livelihoods. To understand this channel, we study the heterogeneity in response to the Covid-19 pandemic by the level of its negative impact on the local economy. We define an area to be more (less) susceptible to the negative impact if the proportion of the area's annual average employment in "Leisure and Hospitality" and "Education and Health Services", the two hardest-hit industries in employment according to Bureau of Labor Statistics, is above (below) the sample median (i.e., 32 percent in the Google sample and 35 percent in the Twitter sample). Figure A9

---

because a video tweet unrelated to Covid-19 but with n-word in the description went extremely viral on February 10th and contributed to 95 percent of the n-word tweets on that day.

[19] We define a tweet to be related to China if it contains any of "China", "Chinese", "Huawei", or "Xi" and a tweet to be related to Covid-19 if it contains any of "covid", "covid-19", "corona", "coronavirus", "virus", "epidemic", or "pandemic". Table A5 presents examples of Trump's tweets in each category, and Figure A8 plots the daily frequency of such tweets. We only include data after January 1st 2020 because Trump did not tweet about Covid-19 until late January 2020.

shows areas experiencing high versus low negative economic impact do not respond differently to the first local Covid-19 diagnosis. One potential reason is that the impact of the pandemic on local economy was not well understood at the onset of the first Covid-19 diagnosis. According to surveys administered in early March and April 2020 (Binder, 2020; Saad, 2020), Americans were more worried about the effect of Covid-19 on their health than on their personal finances.

Taken together, our findings imply that the pressing infection risk and the salience of the connection between the disease and the Asian population play a bigger part than future economic impact in motivating the initial racial animus.

## 4. Conclusion

In this paper, we estimate the effect of the Covid-19 pandemic on racial animus, as measured by Google searches and Twitter posts including the c-word. Our main finding is that the Covid-19 pandemic sparks racial animus against Asians. Shortly after the first Covid-19 diagnosis in an area, the locals start to search for and tweet the c-word more often. Fear of contracting Covid-19 and the salient connection between the disease and the Asian population is a main driving force for the rising animosity against Asians. This increase could indicate a rise in future hate crimes, as we document a strong correlation between the use of the slur and anti-Asian hate crimes using historic data. Moreover, the rise in racially charged tweets mainly comes from existing Twitter users posting the slur for the first time. This can create an multiplier effect on racial animus via persuasion (DellaVigna and Gentzkow, 2010) or by changing the social norms of publicly expressing the anti-Asian sentiment (Bursztyn et al. 2020; Müller and Schwarz, 2019).

Animosity between racial groups could severely hinder global initiatives to tackle the current pandemic and slow economic recovery. Our results suggest that educating the public about the dissemination of Covid-19 and de-emphasizing the connection between the disease and a particular racial group can be an effective way to curb the current and future racial animus.

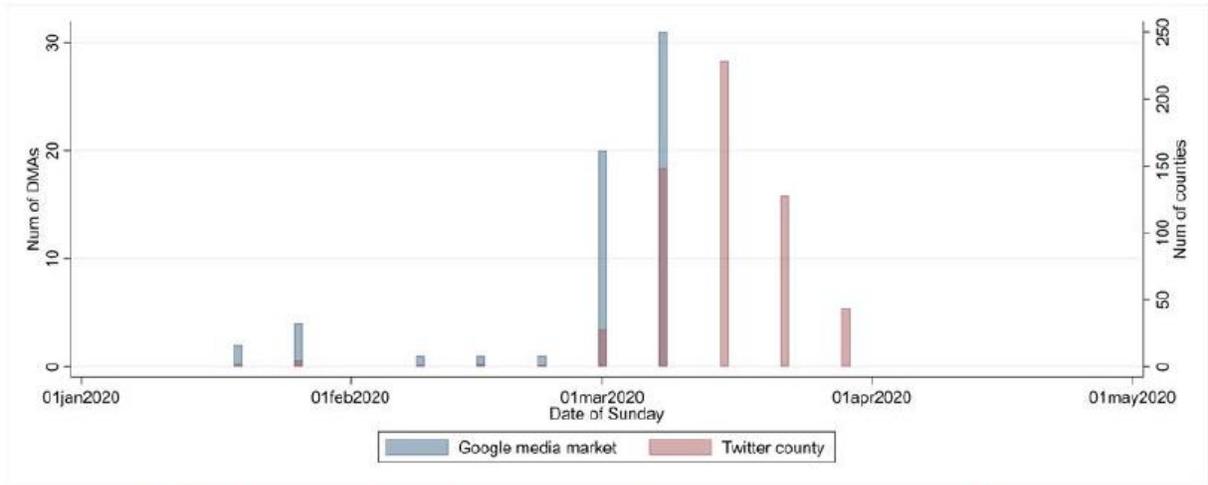

**A:** Number of media markets and counties by week of 1st Covid-19 diagnosis

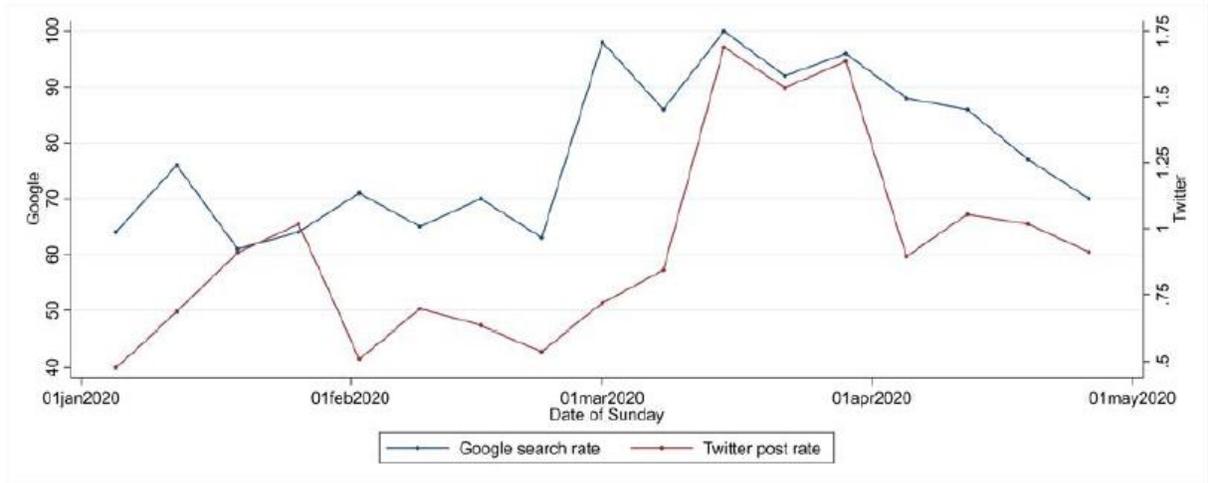

**B:** Racially charged Google search index and Twitter post index

**Figure 1:** Timeline of 1st Local Covid-19 Diagnosis and Use of C-word in the U.S.

*Notes:* Panel A plots the number of media markets and counties in the main regression sample by the week of the first Covid-19 diagnosis in the local area. Panel B plots the weekly Google search index and Twitter post index for "chink(s)" in the United States.

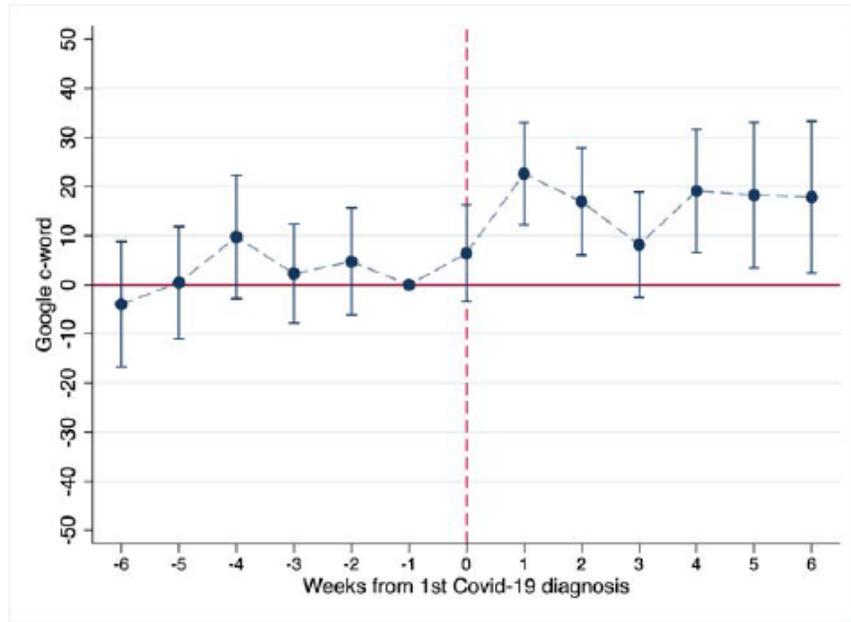

**A:** Google search index

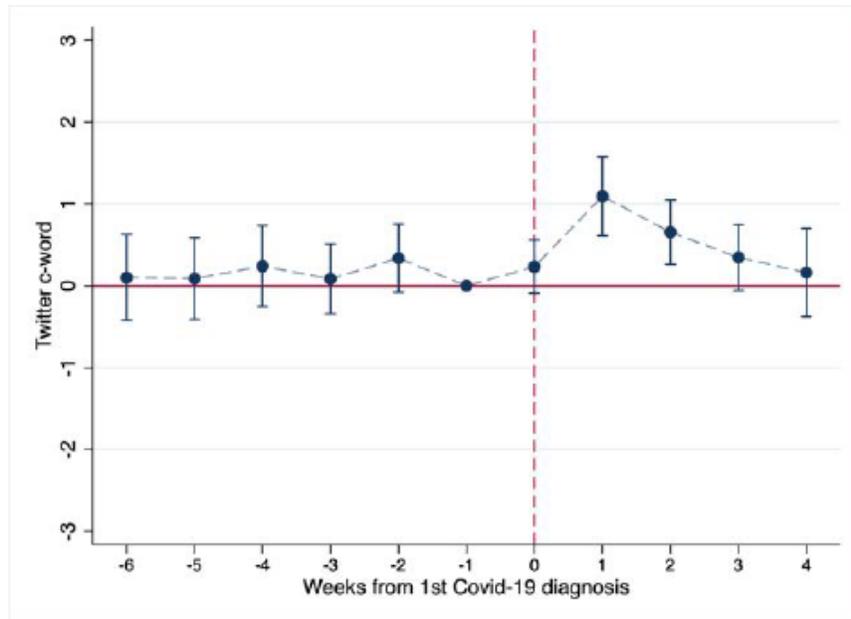

**B:** Twitter post index

**Figure 2:** The Effect of an Area's 1st Covid-19 Diagnosis on Local Racial Animus

*Notes:* The figure presents the effect of the first Covid-19 diagnosis on local racial animus. Panel A and B plot the coefficients and the 95 percent confidence intervals of the event time dummies from regression 2 using racially charged Google search index and racially charged Twitter post index as outcomes, respectively. Regressions control for year-month fixed effects and media market fixed effects (panel A) or year-month fixed effects and county fixed effects (panel B). Standard errors are clustered by media market (panel A) or by county (panel B).

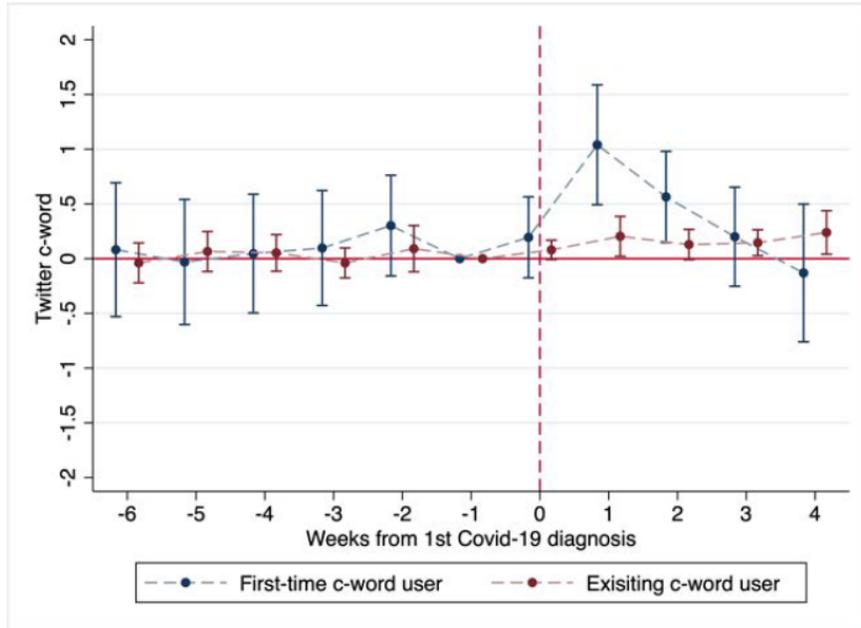

**Figure 3:** The Effect of an Area's 1st Covid-19 Diagnosis on Local Racial Animus by First-time and Existing C-word Users

*Notes:* The figure presents the effect of the first Covid-19 diagnosis on local racial animus from first-time versus pre-existing c-word users. Blue and red lines represent the number of racially charged tweets over the number of "the" tweets from users who have never used the c-word between 2014 and six weeks before their first local Covid-19 diagnosis and had at least 10 other tweets before their first c-word tweet (*first time c-word users*) and those who have used the c-word at least once between 2014 and six weeks before their first local diagnosis (*existing c-word users*), respectively. The coefficients and the 95 percent confidence intervals of the event time dummies come from regression 2 using racially charged Twitter post index as outcome. Regressions control for year-month fixed effects and county fixed effects. Standard errors are clustered by county.

Table 1: Relationship between Local Racial Animus, Hate Crimes, and Chinese Restaurant Visits

| VARIABLES | (1) Incidents/1m | (2) Incidents/1m | (3) Incidents/1m | (4) Incidents/1m | (5) Visits/1m | (6) Visits/1m | (7) Visits/1m | (8) Visits/1m |
|---|---|---|---|---|---|---|---|---|
| **Panel A: Google search index** | | | | | | | | |
| Google c-word(t) | 0.00019** | 0.00019** | 0.00019** | 0.00019** | -5.123 | -5.402* | -6.298* | -6.526** |
|  | (0.00009) | (0.00009) | (0.00009) | (0.00009) | (3.169) | (3.166) | (3.152) | (3.177) |
| Google c-word(t-1) |  |  | -0.00009 | -0.00009 |  |  | -2.773 | -2.917 |
|  |  |  | (0.00009) | (0.00009) |  |  | (2.900) | (2.904) |
| Google Asian(s)(t) |  | -0.00005 |  | -0.00009 |  | -33.567** |  | -21.114** |
|  |  | (0.00030) |  | (0.00038) |  | (15.637) |  | (9.296) |
| Google Asian(s)(t-1) |  |  |  | 0.00014 |  |  |  | -10.274 |
|  |  |  |  | (0.00061) |  |  |  | (10.078) |
| Total visits/1m |  |  |  |  | 0.030*** | 0.030*** | 0.029*** | 0.029*** |
|  |  |  |  |  | (0.007) | (0.007) | (0.005) | (0.006) |
| Observations | 3,600 | 3,600 | 3,600 | 3,600 | 1,440 | 1,440 | 1,380 | 1,380 |
| R-squared | 0.18620 | 0.18621 | 0.18642 | 0.18645 | 0.980 | 0.980 | 0.985 | 0.985 |
| Outcome mean | .03746 | .03746 | .03746 | .03746 | 26353.271 | 26353.271 | 26353.271 | 26353.271 |
| **Panel B: Twitter post index** | | | | | | | | |
| Twitter c-word | -0.00008 | -0.00008 | -0.00008 | -0.00008 | -3.25768 | -6.22705 | -15.11324 | -15.83304 |
|  | (0.00016) | (0.00016) | (0.00015) | (0.00015) | (37.49299) | (36.73437) | (40.16780) | (39.91470) |
| Twitter c-word (t-1) |  |  | -0.00010 | -0.00009 |  |  | -15.36303 | -15.03136 |
|  |  |  | (0.00007) | (0.00007) |  |  | (20.76872) | (20.60958) |
| Twitter Asian(s)(t) |  | -0.00000** |  | -0.00000*** |  | 1.44369 |  | 0.50875 |
|  |  | (0.00000) |  | (0.00000) |  | (0.93082) |  | (0.65215) |
| Twitter Asian(s)(t-1) |  |  |  | -0.00000 |  |  |  | 0.35386 |
|  |  |  |  | (0.00000) |  |  |  | (0.87810) |
| Total visits/1m |  |  |  |  | 0.02591*** | 0.02589*** | 0.02495*** | 0.02493*** |
|  |  |  |  |  | (0.00525) | (0.00523) | (0.00424) | (0.00422) |
| Observations | 12,104 | 12,104 | 11,847 | 11,847 | 4,932 | 4,932 | 4,698 | 4,698 |
| R-squared | 0.04462 | 0.04550 | 0.04517 | 0.04619 | 0.97053 | 0.97056 | 0.97499 | 0.97499 |
| Outcome mean | .0033 | .0033 | .0033 | .0033 | 23708.533 | 23708.533 | 23708.533 | 23708.533 |

*Notes*: The table correlates the racially charged Google search index and Twitter post index with anti-Asian hate crimes and visits to Chinese restaurants. All data are at the media market×year-month level. Outcome variables are the number of anti-Asian hate crimes per one million population between Jan 2014 and Dec 2018 (columns (1)-(4)) and the number of visits to Chinese restaurants per one million population between Jan 2018 and Dec 2019 (columns (5)-(8)). All regressions control for local unemployment rate and year-month fixed effects as well as media market fixed effects. All regressions are weighted by local population. *** p<0.01, ** p<0.05, * p<0.1.

Table 2: Relationship between Local Racial Animus and Trump Tweets

| VARIABLES | (1) Twitter c-word | (2) Twitter c-word | (3) Twitter c-word | (4) Google c-word | (5) Google c-word | (6) Google c-word |
|---|---|---|---|---|---|---|
| China & Covid(t) | 0.0798*** | 0.0829*** | 0.0789*** | 0.4466 | 0.5693* | 0.5029* |
|  | (0.0251) | (0.0250) | (0.0257) | (0.2865) | (0.2938) | (0.2866) |
| China only(t) | -0.0327** | -0.0208 | -0.0189 | 0.2848 | 0.3157 | 0.2996 |
|  | (0.0139) | (0.0145) | (0.0145) | (0.1961) | (0.2141) | (0.2159) |
| Covid only(t) | 0.0160*** | 0.0012 | -0.0029 | 0.0529 | 0.0000 | 0.0122 |
|  | (0.0047) | (0.0052) | (0.0054) | (0.0594) | (0.0745) | (0.0758) |
| New diagnoses |  |  | 0.0000*** |  |  | -0.0001 |
|  |  |  | (0.0000) |  |  | (0.0001) |
| New deaths |  |  | -0.0002** |  |  | -0.0002 |
|  |  |  | (0.0001) |  |  | (0.0010) |
| Twitter Asian(s) |  |  | 0.0011*** |  |  |  |
|  |  |  | (0.0002) |  |  |  |
| Google Asian(s) |  |  |  |  |  | -0.0244 |
|  |  |  |  |  |  | (0.0190) |
| Observations | 28,329 | 28,329 | 28,329 | 7,380 | 7,380 | 7,380 |
| R-squared | 0.0019 | 0.0333 | 0.0378 | 0.0016 | 0.0592 | 0.0596 |
| Outcome mean | .669 | .669 | .669 | 8.037 | 8.037 | 8.037 |
| County FE | N | Y | Y | N | N | N |
| DMA FE | N | N | N | N | Y | Y |
| Ym FE | N | Y | Y | N | Y | Y |
| DOW FE | N | Y | Y | N | Y | Y |

*Notes*: The table correlates the racially charged Google search index and Twitter post index with President Trump's tweets about Covid-19 or China. The data are at the media market×daily level (columns (1)-(3)) or county×daily level (columns (4)-(6)) between Jan 1st 2020 and May 2nd 2020. Outcome variables are racially charged Twitter post index (columns (1)-(3)) and racially charged Google search index (columns (4)-(6)). A tweet is defined to be about China if it contains any of "China", "Chinese", "Huawei", or "Xi" and about Covid-19 if it contains any of "covid", "covid-19", "corona", "coronavirus", "virus", "epidemic", or "pandemic". "New diagnoses" and "New deaths" are the total daily number of new Covid-19 diagnoses and deaths in the United States calculated using the data from John Hopkins University Covronavirus Resource Center. Standard errors in parentheses are clustered by county (columns (1)-(3)) or by media market (columns (4)-(6)). *** p<0.01, ** p<0.05, * p<0.1.

**Appendix A. Construction of Comparable Google Search Index**

Google Trends only reports search index in either a time series format or a cross-sectional format. To construct a panel data set consisting of time series search index in all possible media markets, we need to extract the search index in media market separately. However, the search index reported by Google Trends is not the level of search rate but the search rate normalized by the maximum search rate appearing in that extraction. As a result, search indices from two extractions do not share the same base. To build a panel of search indices that are normalized by the same value, we can scale the search index using the following method.

In a time series extraction of search index in media market $m$ over period $T$, the search index in media market $m$ at time $t$ is approximately:

$$\text{Search Index}_{mt,T} = 100 \times \frac{\frac{\text{Searches including ``chink(s)''}_{mt}}{\text{Total searches}_{mt}}}{\max_{t \in T}\{\frac{\text{Searches including ``chink(s)''}_{mt}}{\text{Total searches}_{mt}}\}} \quad (3)$$

Meanwhile, in a cross-sectional extraction of search index at time $t$ for all media markets $m \in M$, the search index in media market $m$ at time $t$ is approximately:

$$\text{Search Index}_{mt,M} = 100 \times \frac{\frac{\text{Searches including ``chink(s)''}_{mt}}{\text{Total searches}_{mt}}}{\max_{m \in M}\{\frac{\text{Searches including ``chink(s)''}_{mt}}{\text{Total searches}_{mt}}\}} \quad (4)$$

If we assume that the numerators in equations 3 and 4 are the same and both search indices are non-zero, we can calculate the ratio of the two denominators as:

$$Ratio_{m.MT} = \frac{\max_{t \in T}\{\frac{\text{Searches including ``chink(s)''}_{mt}}{\text{Total searches}_{mt}}\}}{\max_{m \in M}\{\frac{\text{Searches including ``chink(s)''}_{mt}}{\text{Total searches}_{mt}}\}} = \frac{\text{Search Index}_{mt,M}}{\text{Search Index}_{mt,T}} \quad (5)$$

We can then scale the time series search index over period $T$ in each media market $m \in M$ by multiplying it with its corresponding $Ratio_{m,MT}$. The resulting time series will be normalized by the same $\max_{m \in M}\{\frac{Searches\ including\ "chink(s)"_{mt}}{Total\ searches_{mt}}\}$.

The practical obstacle for the above re-scaling is that Google trends returns zero if the absolute level of search for a given media market and time is below an unreported threshold. The re-scaling does not work if either $Searches\ Index_{m,tT}$ or $Searches\ Index_{m,tM}$ is zero. After extracting cross-sectional search indices on all possible $t \in T$, we can at best back out the comparable search index for 35 out of 60 media markets normalized by the maximum search rate of Fresno-Visalia media market in California over the sample period, using the cross-sectional search index on March 15th, 2020 with all search index. Given the small sample size, we only report the results using the comparable search index in the appendix.

# Appendix B. Additional Figures & Tables

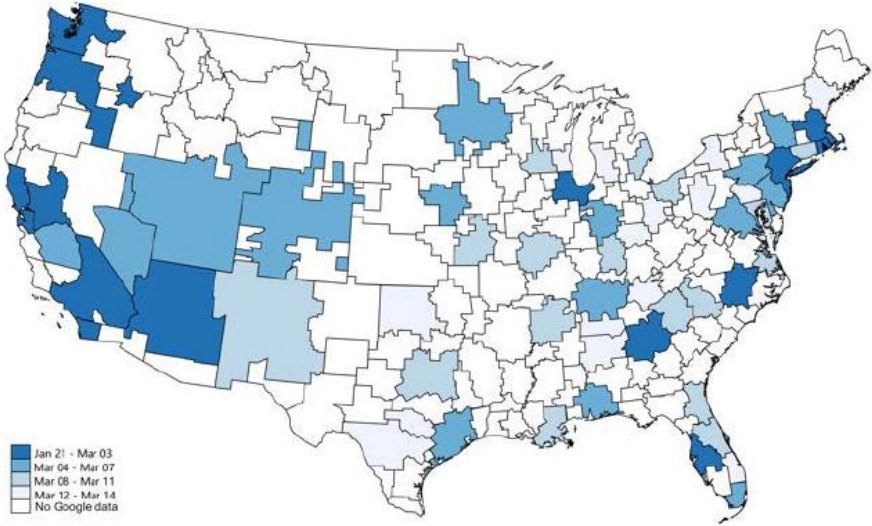

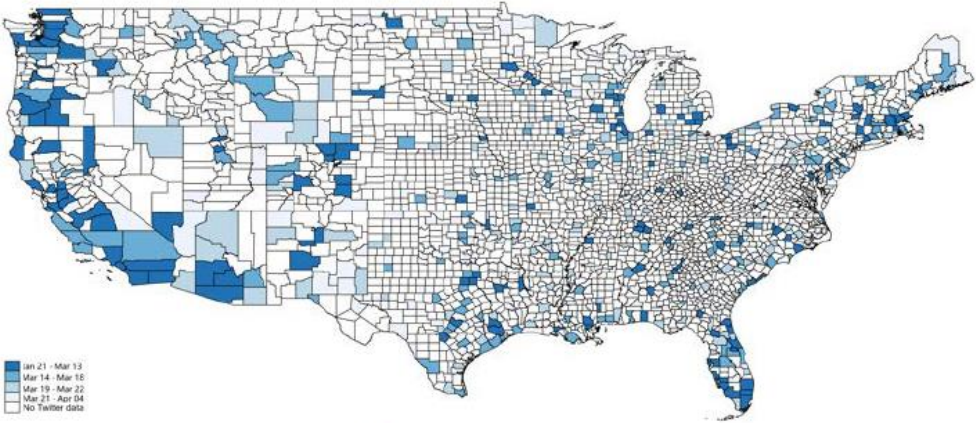

**Figure A1:** Media Markets and Counties by Timing of 1st Covid-19 Diagnosis

*Notes:* The figure presents the map of the media markets (panel A) and the counties (panel B) in the main regression sample by the date of the first Covid-19 diagnosis in the local area. The darker the color, the earlier the first diagnosis.

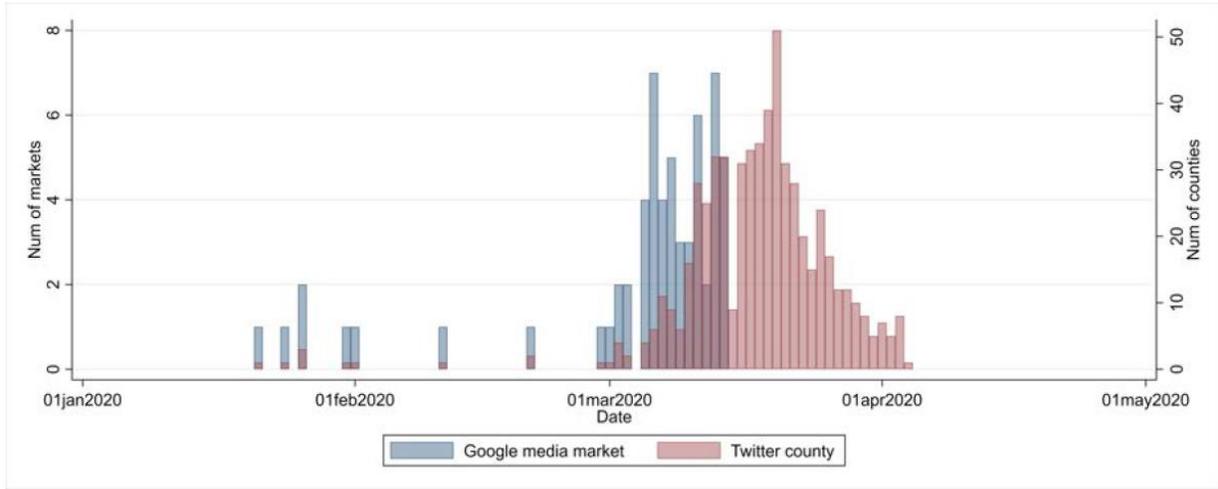

**Figure A2:** Number of Media Markets and Counties by Day of 1st Covid-19 Diagnosis

*Notes:* This figure plots the number of the media markets (blue bar) and the number of counties (red bar) in the main regression sample by the date of the first Covid-19 diagnosis in the local area.

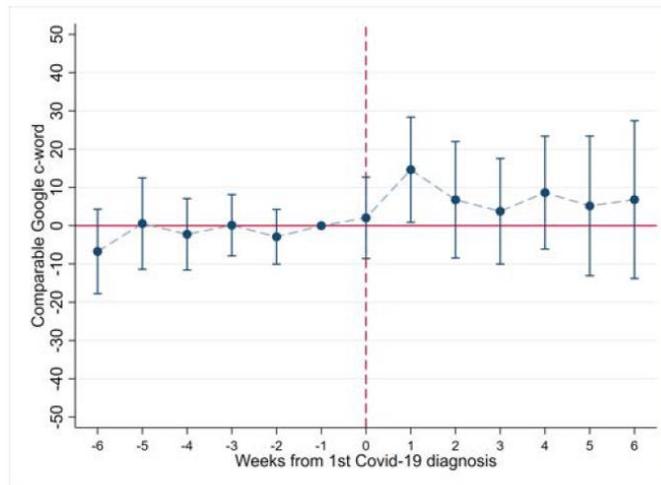

**Figure A3:** Robustness - The Effect of an Area's 1st Covid-19 Diagnosis on Local Racial Animus using Re-scaled Google Index

*Notes:* The figure presents the effect of the first Covid-19 diagnosis on local racial animus. It plots the coefficients and the 95 percent confidence intervals of the event time dummies from regression 2 using re-scaled racially charged Google search index. Regressions control for year-month fixed effects and media market fixed effects. Standard errors are clustered by media market.

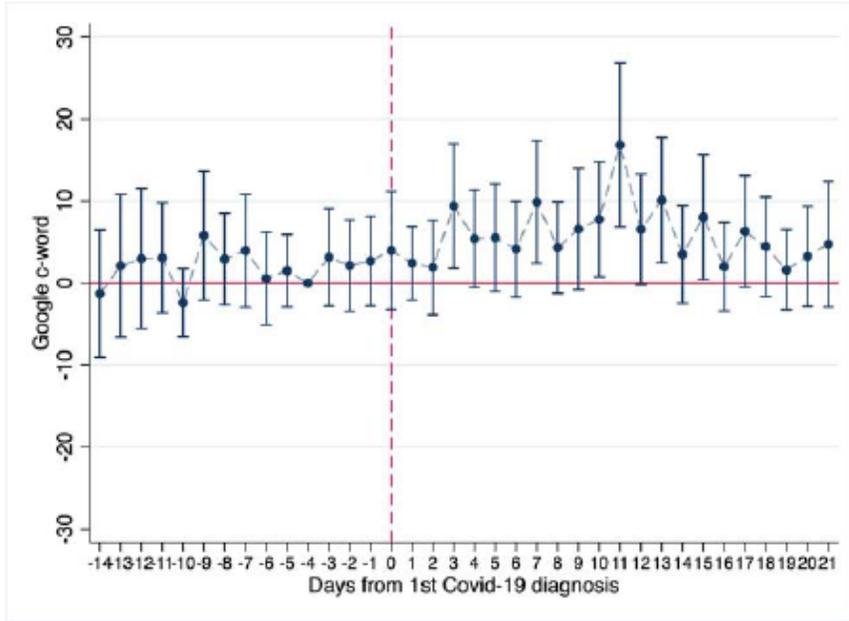

**A:** Google search rate

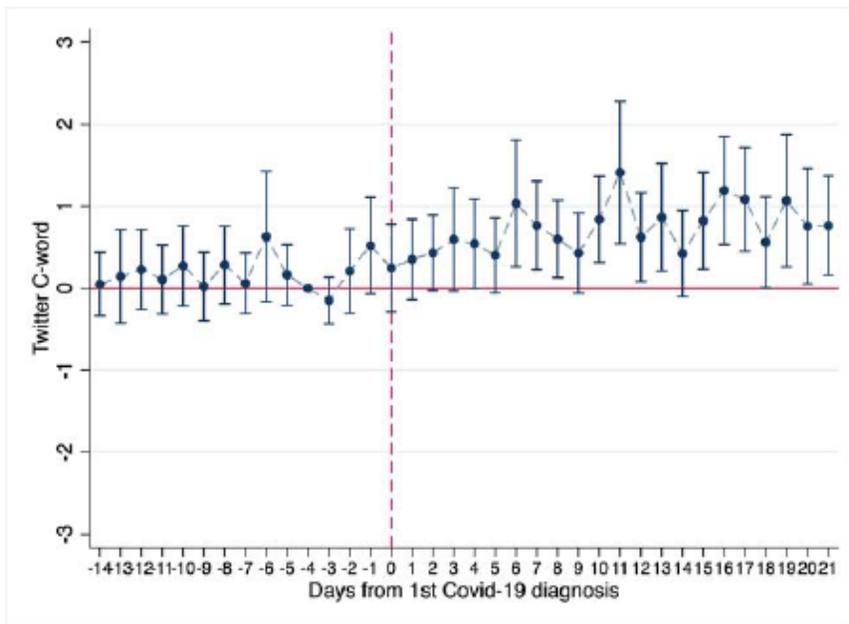

**B:** Twitter post rate

**Figure A4:** The Effect of an Area's 1st Covid-19 Diagnosis on Daily Racial Animus

*Notes:* The figures present the effect of the first Covid-19 diagnosis on local racial animus at the daily level. Panel A and B plot the coefficients and the 95 percent confidence intervals on the event time dummies from 14 days before to 21 days after the day of the first Covid-19 diagnosis using racially charged Google search index and racially charged Twitter post index as outcomes, respectively. Regressions control for year-month fixed effects, day-of-week fixed effects, and media market fixed effects (panel A) or county fixed effects (panel B). Standard errors are clustered by media market (panel A) or by county (panel B).

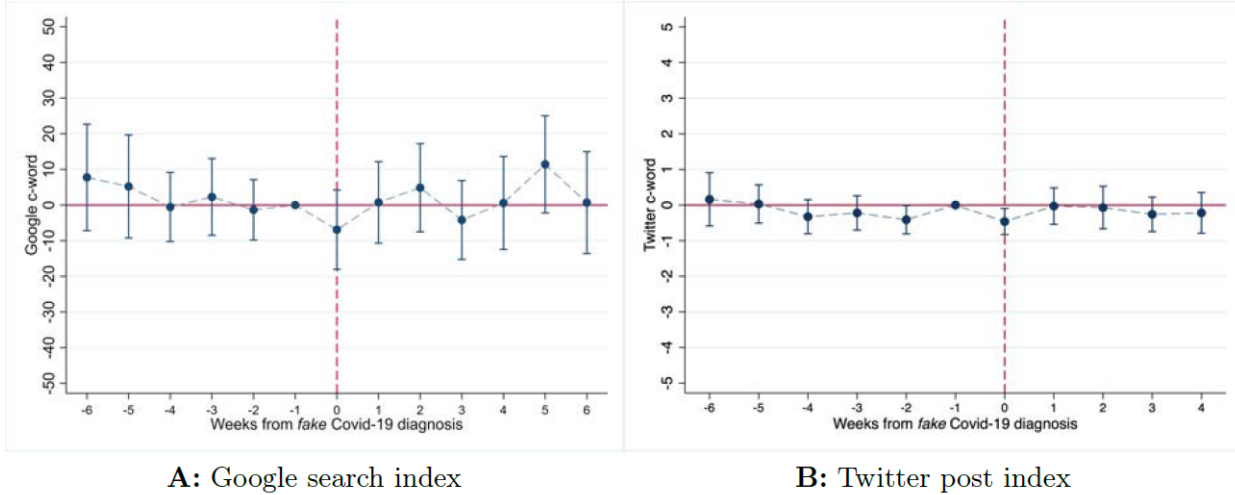

**A:** Google search index        **B:** Twitter post index

**Figure A5:** Placebo - The Effect of an Area's *fake* Covid-19 Diagnosis on Local Racial Animus

*Notes:* The figure presents a placebo test for the effect of the first Covid-19 diagnosis on local racial animus. We replace the date of the first local Covid-19 diagnosis with a *fake* date which shares the same day and month as the actual date but in year 2019 instead of 2020. Panel A and B plot the coefficients and the 95 percent confidence intervals of the event time dummies based on the *fake* date from regression 2 using racially charged Google search index and racially charged Twitter post index as outcomes, respectively. Regressions control for year-month fixed effects and media market fixed effects (panel A) or year-month fixed effects and county fixed effects (panel B). Standard errors are clustered by media market (panel A) or county (panel B).

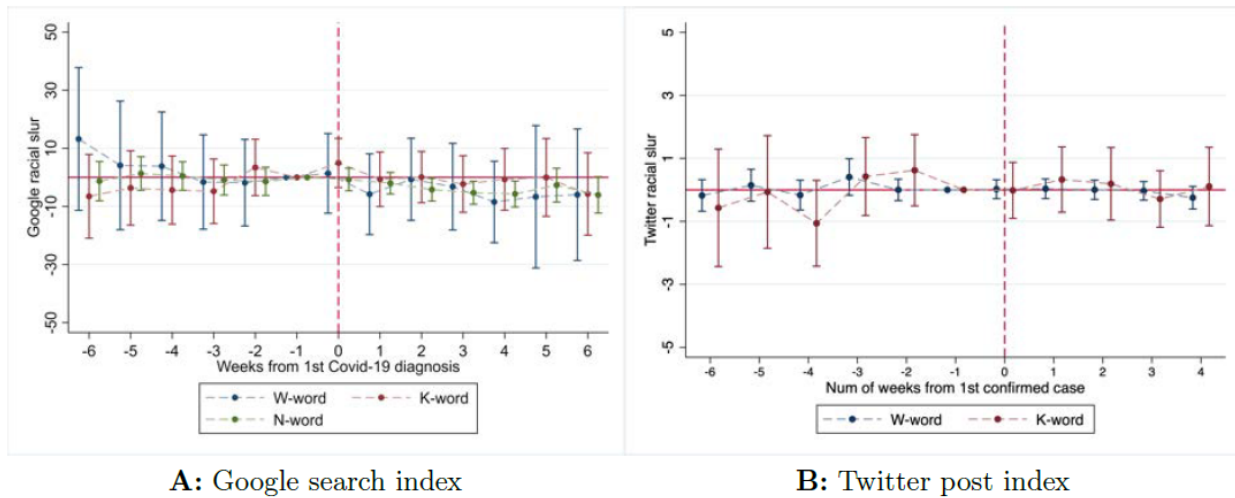

**A:** Google search index        **B:** Twitter post index

**Figure A6:** The Effect of an Area's 1st Covid-19 Diagnosis on Local Racial Animus Against Other Minority Groups

*Notes:* The figure presents the effect of the first Covid-19 diagnosis on local racial animus against three other minority groups - Hispanics, Jews, and African Americans, proxy by Google search index and Twitter post index of "wetback(s)", "kike(s)", and "nigger(s)" respectively. Regression sample for n-word, k-word, and w-word Google search index contains 203, 78, and 27 media markets (panel A). Regression sample for w-word and k-word Twitter post index contain 587 counties (panel B). The displayed coefficients and the 95 percent confidence intervals of the event time dummies are from regression 2 using racially charged Google search index and Twitter post index as outcomes. All regressions control for year-month fixed effects and media market fixed effects (panel A) or year-month fixed effects and county fixed effects (panel B). We include an indicator for the week of January 26th in the regression for n-word to control for the spike due to Kobe Bryant's death. We also include an indicator for the week of February 23th in the regression for k-word to control for the spike due to Los Angeles Dodges player Enrique ("Kiké") Hernandez's performance in that week. Standard errors are clustered by media market (panel A) or by county (panel B).

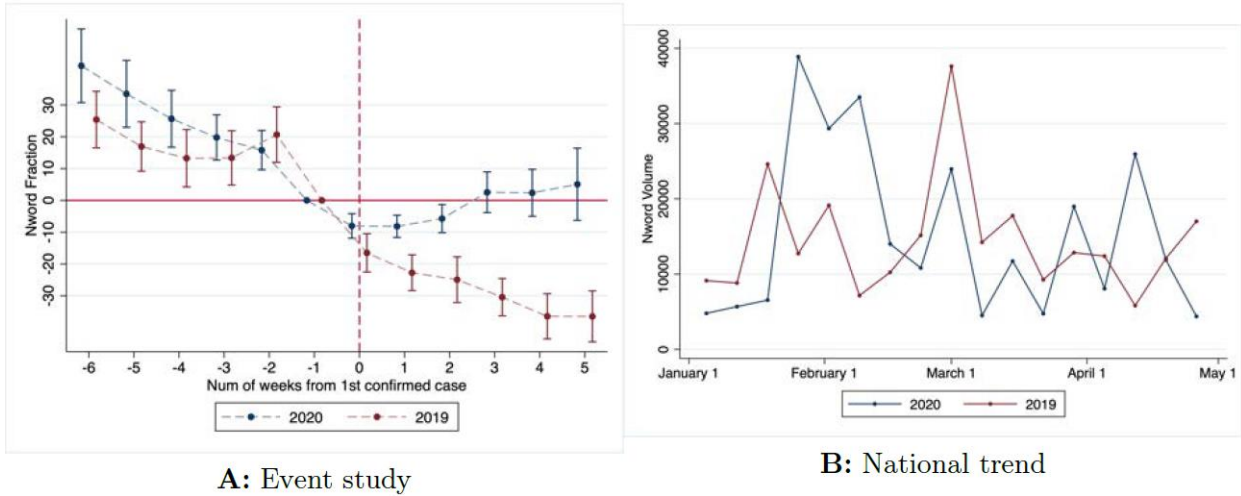

**A:** Event study  **B:** National trend

**Figure A7:** Racial Animus Against African Americans on Twitter

*Notes:* Figure A presents estimates of the coefficients on the event dummies from regression 2, using Twitter post index of the n-word between November 2019 and April 2020 (blue line) and that between November 2018 and April 2019 (red line) as the outcomes. For the regression using 2018-2019 data, we replace the date of the first local Covid-19 diagnosis with a *fake* date which shares the same day and month as the actual date in 2020 but with the year as 2019. For the regression using 2019-2020 data, we include an indicator for the week of January 26th (Kobe Bryant's death) and another for the week of February 9th (a video tweet unrelated to Covid-19 but with n-word in the description went extremely viral on Twitter on February 10th). Figure B presents the national time trend for the Twitter post index of the n-word in 2020 (blue line) and 2019 (red line).

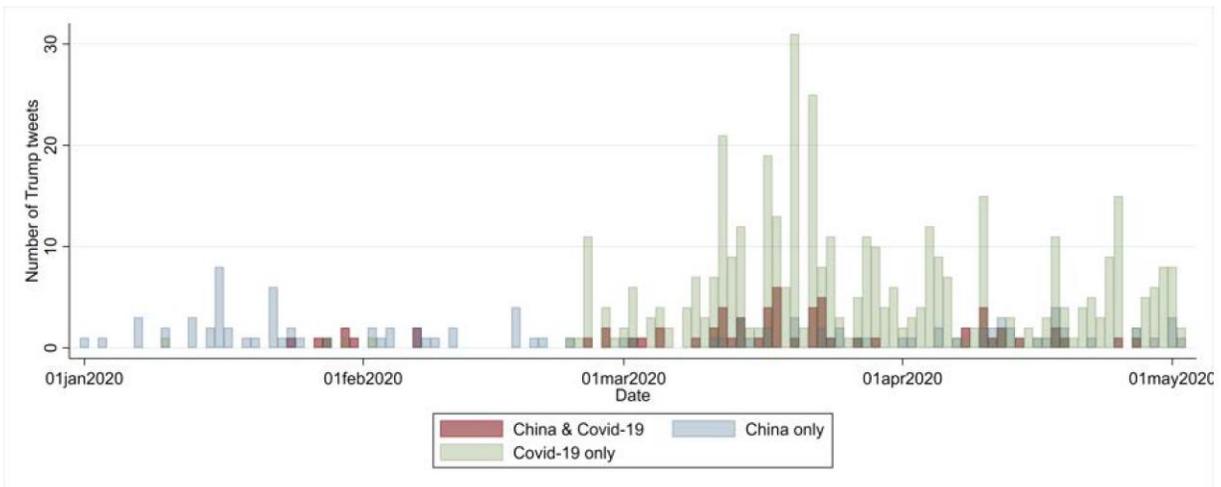

**Figure A8:** Number of President Trump's Tweets about China or Covid-19

*Notes:* This figure plots the daily number of President Trump's tweets on both China and Covid-19 (blue bar), only China (red bar), and only Covid-19 (green bar) between Jan 1st 2020 and May 2nd 2020. A tweet is defined to be about China if it contains any of "China", "Chinese", "Huawei", or "Xi" and about Covid-19 if it contains any of "covid", "covid-19", "corona", "coronavirus", "virus", "epidemic", or "pandemic".

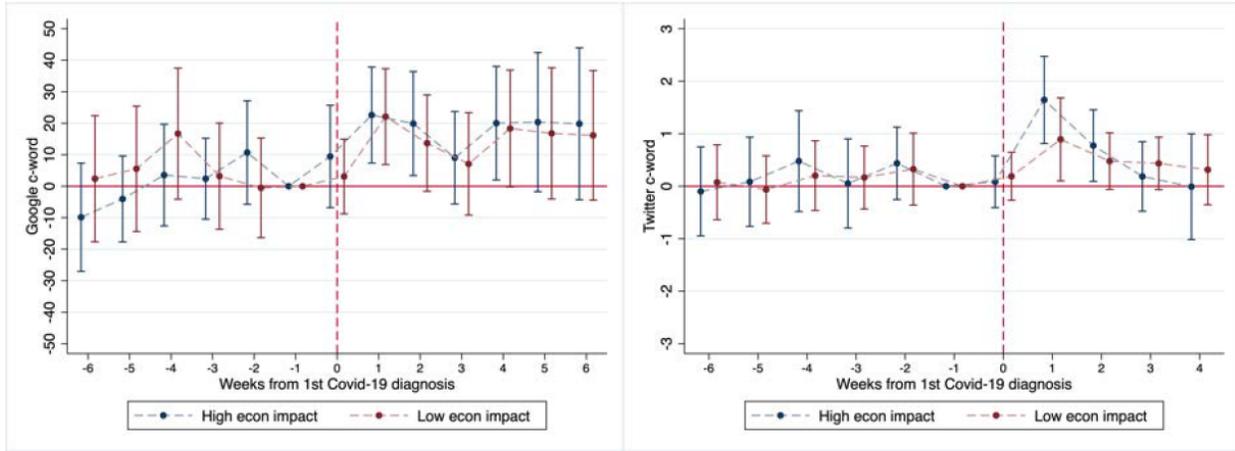

**A:** Google search index          **B:** Twitter post index

**Figure A9:** The Effect of an Area's 1st Covid-19 Diagnosis on Local Racial Animus by Negative Economic Impact Caused by the Covid-19 pandemic

*Notes:* The figures present the heterogeneous effect of Covid-19 diagnosis by the local labor market's susceptibility to unemployment caused by the Covid-19 pandemic. A media market or county is defined to endure high (low) economic impact if the annual average proportion of the local employment in the top two hardest-hit industries by the Covid-19 pandemic, i.e., 'Leisure and Hospitality" and "Education and Health Services", is above (below) the sample median (32 percent in panel A and 35 percent in panel B). The employment data is the county level QCEW NAICS-Based Data in 2018 from Bureau of Labor Statistics. Panel A and B plot the coefficients and the 95 percent confidence intervals of the event time dummies from regression 2 using racially charged Google search index and racially charged Twitter post index as outcomes, respectively. Regressions control for year-month fixed effects and media market fixed effects (panel A) or year-month fixed effects and county fixed effects (panel B). Standard errors are clustered by media market (panel A) or county (panel B).

**Table A1:** Sample Selection - Media Markets and Counties with Google and Twitter data

| VARIABLES | (1) Google sample | (2) Google sample | (3) Twitter data | (4) Twitter data | (5) Twitter sample | (6) Twitter sample |
|---|---|---|---|---|---|---|
| Log(pop) | 0.241*** | 0.224*** | 0.120*** | 0.145*** | 0.122*** | 0.145*** |
|  | (0.029) | (0.035) | (0.006) | (0.007) | (0.006) | (0.007) |
| % Asian | 0.025 | 0.060* | 0.005 | 0.009 | 0.006 | 0.009 |
|  | (0.019) | (0.031) | (0.008) | (0.009) | (0.008) | (0.009) |
| % Asian$^2$ | -0.001* | -0.002* | -0.000 | -0.001** | -0.000 | -0.001** |
|  | (0.000) | (0.001) | (0.000) | (0.000) | (0.000) | (0.000) |
| % Male | -0.016 | -0.007 | 0.003 | -0.002 | 0.002 | -0.002 |
|  | (0.026) | (0.040) | (0.002) | (0.002) | (0.002) | (0.002) |
| % 65+ | 0.002 | -0.010 | -0.002 | -0.001 | -0.002 | -0.001 |
|  | (0.009) | (0.016) | (0.002) | (0.002) | (0.001) | (0.002) |
| % BA+ | 0.012*** | 0.007 | 0.003*** | 0.002** | 0.003** | 0.002* |
|  | (0.004) | (0.007) | (0.001) | (0.001) | (0.001) | (0.001) |
| % Unemp | 0.002 | -0.005 | -0.007* | 0.000 | -0.009** | -0.002 |
|  | (0.011) | (0.020) | (0.004) | (0.005) | (0.004) | (0.005) |
| % VS dem-rep | -0.001 | -0.001 | 0.001*** | 0.002*** | 0.001*** | 0.001*** |
|  | (0.001) | (0.002) | (0.000) | (0.000) | (0.000) | (0.000) |
|  |  |  |  |  |  |  |
| Observations | 205 | 205 | 3,111 | 3,111 | 3,111 | 3,111 |
| R-squared | 0.581 | 0.678 | 0.279 | 0.351 | 0.292 | 0.357 |
| Outcome mean | .292 | .292 | .193 | .193 | .186 | .186 |
| State FE | N | Y | N | Y | N | Y |

*Notes*: The table presents the sample selection in Google and Twitter data. The data are at the media market level (columns (1)-(2)) or county level (columns (3)-(6)). Outcome is an indicator of having valid racially charged Google search index (columns (1)-(2)), an indicator of having valid racially charged Twitter post index (columns (3)-(4)), or an indicator of being in the final Twitter sample. Note that all media markets with valid Google data are in the final Google sample. "%Asian", "% Male", "% 65+", and "% BA+" are the percentage of Asians, males, population 65 years old or over, and population with Bachelor's or above degree in the local area from American Community Survey 2014-2018 five-year average. "%Unemp" is the average monthly local unemployment rate between 2014 and 2018 from Bureau of Labor Statistics. "Log(pop)" is the logarithm of local population estimates in 2018 from Census Bureau. "% Vote share dem-rep" is the difference between the democratic and the republican vote share in 2012 presidential election from Havard Dataverse. The number of media markets and counties is less than 210 and 3141 due to missing covariates. Standard errors in parentheses are clustered by media market (columns (1)-(2)) or by county (columns (3)-(4)). *** p<0.01, ** p<0.05, * p<0.1.

**Table A2:** Sample Selection - Timing of 1st Local Covid-19 Diagnosis

| VARIABLES | (1) Google sample Weeks from Jan192020 | (2) Google sample Weeks from Jan192020 | (3) Twitter sample Weeks from Jan192020 | (4) Twitter sample Weeks from Jan192020 |
|---|---|---|---|---|
| Log(pop) | -0.838* | -0.835*** | -0.586*** | -0.575*** |
|  | (0.469) | (0.291) | (0.048) | (0.059) |
| % Asian | -0.184 | 0.004 | -0.058 | -0.068 |
|  | (0.166) | (0.184) | (0.039) | (0.041) |
| % Asian$^2$ | 0.002 | -0.000 | 0.000 | -0.000 |
|  | (0.005) | (0.004) | (0.001) | (0.001) |
| % Male | -0.841** | -0.943** | -0.036 | -0.010 |
|  | (0.391) | (0.364) | (0.028) | (0.029) |
| % 65+ | -0.128 | -0.056 | -0.014 | -0.013 |
|  | (0.077) | (0.048) | (0.009) | (0.014) |
| % BA+ | -0.039 | 0.040 | -0.016*** | -0.005 |
|  | (0.053) | (0.041) | (0.005) | (0.006) |
| % Unemp | -0.219 | 0.477** | -0.013 | 0.047 |
|  | (0.254) | (0.196) | (0.025) | (0.045) |
| % VS dem-rep | 0.002 | -0.019* | -0.001 | -0.002 |
|  | (0.009) | (0.009) | (0.002) | (0.002) |
| Observations | 60 | 60 | 581 | 581 |
| R-squared | 0.529 | 0.975 | 0.510 | 0.600 |
| Outcome mean | 5.983 | 5.983 | 7.913 | 7.913 |
| State FE | N | Y | N | Y |

*Notes*: The table presents the relationship between the timing of first local Covid-19 diagnosis and characteristics of the local area. The data are at the media market level (columns (1)-(2)) or county level (columns (3)-(4)). Outcome is the number of weeks from the week of the first diagnosis in our sample, i.e., the week of Jan 19th 2020. "%Asian", "% Male", "% 65+", and "% BA+" are the percentage of Asians, males, population 65 years old or over, and population with Bachelor's or above degree in the local area from the 2014-2018 five-year average of the American Community Survey. "%Unemp" is the average monthly unemployment rate between 2014 and 2018 from Bureau of Labor Statistics. "Log(pop)" is the logarithm of local population estimates in 2018 from Census Bureau. "% Vote share dem-rep" is the percentage difference in the democratic and the republican vote share in 2012 presidential election from Havard Dataverse. The number of observations in columns (3) and (4) are smaller than that in Table A4 due to missing covariates. Standard errors in parentheses are clustered by media market (columns (1)-(2)) or by county (columns (3)-(4)). *** p<0.01, ** p<0.05, * p<0.1.

**Table A3:** The Effect of an Area's 1st Covid-19 Diagnosis on Local Racial Animus Google Search Index

| VARIABLES | (1) C-word index | (2) Severity control | (3) Asian control | (4) Exclude states |
|---|---|---|---|---|
| -6w | -3.920 | -2.694 | -4.265 | -8.979 |
|  | (6.379) | (6.620) | (6.404) | (8.341) |
| -5w | 0.431 | 1.100 | -0.198 | -2.575 |
|  | (5.722) | (5.820) | (5.699) | (7.083) |
| -4w | 9.764 | 10.088 | 9.419 | 9.205 |
|  | (6.263) | (6.316) | (6.233) | (7.649) |
| -3w | 2.282 | 2.503 | 2.247 | 2.458 |
|  | (5.023) | (5.085) | (5.020) | (5.912) |
| -2w | 4.739 | 4.899 | 4.771 | 2.564 |
|  | (5.469) | (5.535) | (5.467) | (6.150) |
| +0w | 6.421 | 6.326 | 6.274 | 6.574 |
|  | (4.898) | (4.911) | (4.864) | (5.127) |
| +1w | 22.628*** | 22.442*** | 22.030*** | 22.771*** |
|  | (5.210) | (5.246) | (5.280) | (5.721) |
| +2w | 16.945*** | 15.936*** | 16.727*** | 18.104*** |
|  | (5.439) | (5.443) | (5.407) | (5.621) |
| +3w | 8.155 | 5.702 | 7.894 | 8.614 |
|  | (5.359) | (5.907) | (5.403) | (5.829) |
| +4w | 19.106*** | 15.972** | 18.873*** | 19.527** |
|  | (6.265) | (6.999) | (6.253) | (7.461) |
| +5w | 18.263** | 15.375* | 18.041** | 14.709* |
|  | (7.411) | (8.113) | (7.428) | (8.679) |
| +6w | 17.861** | 15.002* | 18.125** | 18.017* |
|  | (7.726) | (8.046) | (7.751) | (9.267) |
| New cases(t) |  | 0.000 |  |  |
|  |  | (0.000) |  |  |
| New deaths(t) |  | -0.006 |  |  |
|  |  | (0.004) |  |  |
| Post lockdown |  | 3.691 |  |  |
|  |  | (4.072) |  |  |
|  |  |  |  |  |
| Observations | 780 | 780 | 780 | 663 |
| R-squared | 0.190 | 0.192 | 0.193 | 0.180 |
| Outcome mean | 30.03 | 30.03 | 30.03 | 30.03 |
| Outcome sd | 28.501 | 28.501 | 28.501 | 28.501 |

*Notes:* The table presents the effect of the first Covid-19 diagnosis on local racial animus. All estimates are from regression 2 using racially charged Google search index as outcome. Event dummy for the week before the first local diagnosis is omitted. The first column corresponds to Figure A4 panel A. The second column additionally controls for the number of new case and new deaths and whether the state has stay-at-home orders in place. The third column controls for Google search index for "Asian(s)". The fourth column excludes early- and hard-hit states, namely Washington, New York, and California. All regressions control for media market fixed effects and year-month fixed effects. Standard errors are clustered by media market. *** $p<0.01$, ** $p<0.05$, * $p<0.1$.

**Table A4:** The Effect of an Area's 1st Covid-19 Diagnosis on Local Racial Animus Twitter Post Index

| VARIABLES | (1) C-word index | (2) Severity control | (3) AA control | (4) Exclude states |
|---|---|---|---|---|
| -6w | 0.101 | 0.103 | 0.103 | 0.222 |
|  | (0.267) | (0.268) | (0.267) | (0.285) |
| -5w | 0.089 | 0.090 | 0.086 | 0.173 |
|  | (0.254) | (0.254) | (0.254) | (0.267) |
| -4w | 0.239 | 0.240 | 0.242 | 0.317 |
|  | (0.252) | (0.251) | (0.251) | (0.265) |
| -3w | 0.086 | 0.086 | 0.100 | 0.175 |
|  | (0.218) | (0.219) | (0.219) | (0.240) |
| -2w | 0.337 | 0.337 | 0.336 | 0.360 |
|  | (0.213) | (0.213) | (0.212) | (0.233) |
| +0w | 0.233 | 0.228 | 0.137 | 0.264 |
|  | (0.166) | (0.163) | (0.171) | (0.185) |
| +1w | 1.094*** | 1.076*** | 1.045*** | 0.937*** |
|  | (0.246) | (0.236) | (0.238) | (0.242) |
| +2w | 0.655*** | 0.610*** | 0.754*** | 0.589*** |
|  | (0.201) | (0.219) | (0.202) | (0.221) |
| +3w | 0.346* | 0.272 | 0.497** | 0.388* |
|  | (0.205) | (0.234) | (0.211) | (0.230) |
| +4w | 0.162 | 0.076 | 0.336 | 0.182 |
|  | (0.276) | (0.305) | (0.281) | (0.307) |
| New cases(t) |  | 0.000 |  |  |
|  |  | (0.000) |  |  |
| New deaths(t) |  | -0.001 |  |  |
|  |  | (0.001) |  |  |
| Post lockdown |  | 0.065 |  |  |
|  |  | (0.199) |  |  |
| Observations | 4,796 | 4,796 | 4,796 | 4,251 |
| R-squared | 0.140 | 0.140 | 0.155 | 0.144 |
| Outcome mean | .994 | .994 | .994 | .994 |
| Outcome sd | 3.189 | 3.189 | 3.189 | 3.189 |

*Notes*: The table presents the effect of the first Covid-19 diagnosis on local racial animus. All estimates are from regression 2 using racially charged Twitter post index as outcome. Event dummy for the week before the first local diagnosis is omitted. The first column corresponds to Figure A4 Panel B. The second column additionally controls for the number of new case and new deaths and whether the state has stay-at-home orders in place. The third column controls for Twitter post index for "Asian(s)". The fourth column excludes early- and hard-hit states, namely Washington, New York, and California. All regressions control for county fixed effects and year-month fixed effects. Standard errors are clustered by county. *** p<0.01, ** p<0.05, * p<0.1.

**Table A5:** Examples of President Trump's Tweets about China or Covid-19

| Category | Post | Date |
|---|---|---|
| Only China | "Years from now, when we look back at this day, nobody's going to remember nancy's cheap theatrics, they will remember though how president trump brought the Chinese to the bargaining table and delivered achievements few ever thought were possible." @ingrahamangle @foxnews | 1/17/20 |
| Only China | The Wall Street Journal editorial board doesn't have a clue on how to fight and win. Their views on tariffs & trade are losers for the U.S., but winners for other countries, including China. if we followed their standards, we'd have no country left. They should love sleepy joe! | 4/11/20 |
| Only Covid-19 | The coronavirus is very much under control in the USA. we are in contact with everyone and all relevant countries. CDC & World Health have been working hard and very smart. Stock market starting to look very good to me! | 2/24/20 |
| Only Covid-19 | I am fully prepared to use the full power of the federal government to deal with our current challenge of the coronavirus! | 3/11/20 |
| Covid-19 & China | Just received a briefing on the coronavirus in china from all of our great agencies, who are also working closely with china. we will continue to monitor the ongoing developments. we have the best experts anywhere in the world, and they are on top of it 24/7! | 1/30/20 |
| Covid-19 & China | I will be having a news conference today to discuss very important news from the FDA concerning the Chinese virus! | 3/18/20 |

*Notes*: This table presents examples of President Trump's tweets mentioning China or Covid-19. We define a tweet to be related to China if it contains any of "China", "Chinese", "Huawei", or "Xi" and a tweet to be related to Covid-19 if it contains any of "covid", "covid-19", "corona", "coronavirus", "virus", "epidemic", or "pandemic".